\documentclass[APS, twocolumn]{USG}
\usepackage{anyfontsize} %

\articletype{Research Article}%

\received{00}
\revised{00}
\accepted{00}
\journal{Laser \& Photonics Reviews}
\volume{00}
\copyyear{2025}
\startpage{1}
\articledoi{10.1002/lpor.202501375}

\usepackage{amsmath}
\usepackage{amsfonts}
\usepackage{amssymb}
\usepackage{graphicx}
\usepackage{color}
\usepackage[dvipsnames]{xcolor}
\usepackage{xspace}
\usepackage[english]{babel}
\usepackage{multirow}
\usepackage{xfrac}
\usepackage{nicefrac}
\usepackage{sidecap}

\newcommand{\ud}[1]{{#1^{\dagger}}}

\newcommand\Tr{\mathrm{Tr}}
\newcommand{\mean}[1]{\langle#1\rangle}

\graphicspath{{./images/}}

\begin{document} 

\title{Pulsed Quantum Excitation}

\author[1,2]{Juan Camilo L\'opez~Carre{\~n}o}[https://orcid.org/0000-0002-2134-0958]

\authormark{J.~C. López Carreño}
\titlemark{Pulsed Quantum Excitation}

\address[1]{\orgdiv{Center for Theoretical Physics}, \orgname{Polish
    Academy of Sciences}, \orgaddress{Aleja Lotników 32/46, 02-668
    Warsaw, \country{Poland}}}

\address[2]{\orgdiv{Institute of Experimental Physics, Faculty of
    Physics}, \orgname{University of Warsaw},
  \orgaddress{ul. Pasteura 5, 02-093 Warsaw, \country{Poland}}}

\corres{J. C. López Carreño  (\email{juclopezca@cft.edu.pl})}

\abstract[Abstract]{A two-level system (2LS) is the most fundamental
  building block of matter. Its response to classical light is well
  known, as it converts pulses of coherent light into antibunched
  emission. However, recent theoretical proposals have predicted
  that it is advantageous to illuminate two-level systems with
  \emph{Quantum Light}; i.e., the light emitted from a quantum
  system. However, those proposals were done considering continuous
  excitation of the source of light. Here, we advance the field by
  changing the paradigm of excitation: we use the emission of a 2LS,
  itself driven by a laser pulse, to excite another 2LS. Thus, we
  present a thorough analysis of Resonance Fluorescence under pulsed
  quantum excitation and show, in particular, that the emission from
  a 2LS driven with quantum light is more antibunched and more
  indistinguishable than if it were driven with classical light. Our
  results reinforce the claim of the advantage of the excitation
  with quantum light, provide support to the recent experimental
  observations, and can be used as a road-map for the future of
  light-matter interaction research.}



\maketitle

\section{Introduction}

Excitation with \emph{Quantum Light}, understood as using the light
emitted from a quantum system as a resource to drive optical
targets, is an idea as old as the field of Quantum
Optics~\cite{carmichael1976, yurke1984, collett1984}. The advent of
reliable sources of antibunched~\cite{rempe1991} and squeezed
light~\cite{polzik1992, ou1992} motivated the development of a
theoretical framework to describe adequately such an
excitation. Thus, in 1993 Gardiner~\cite{gardiner1993} and
Carmichael~\cite{carmichael1993} independently introduced the
formalism that we now know as the \emph{theory of cascaded systems},
which is based on the input-output theory of quantum optics. Since
its development, this theory has been used to explore fundamental
topics, including chirality in non-hermitian
systems~\cite{downing2019, sun2023}, the shaping of flying
qubits~\cite{tissot2024}, quantum steering~\cite{xiang2022}; but
also to implement practical applications, such as feeding forward
information into quantum reservoir networks~\cite{ghosh2019a,
  ghosh2020a, dudas2023}.

\begin{figure}[t]
  \begin{center}
    \includegraphics[width=\linewidth]{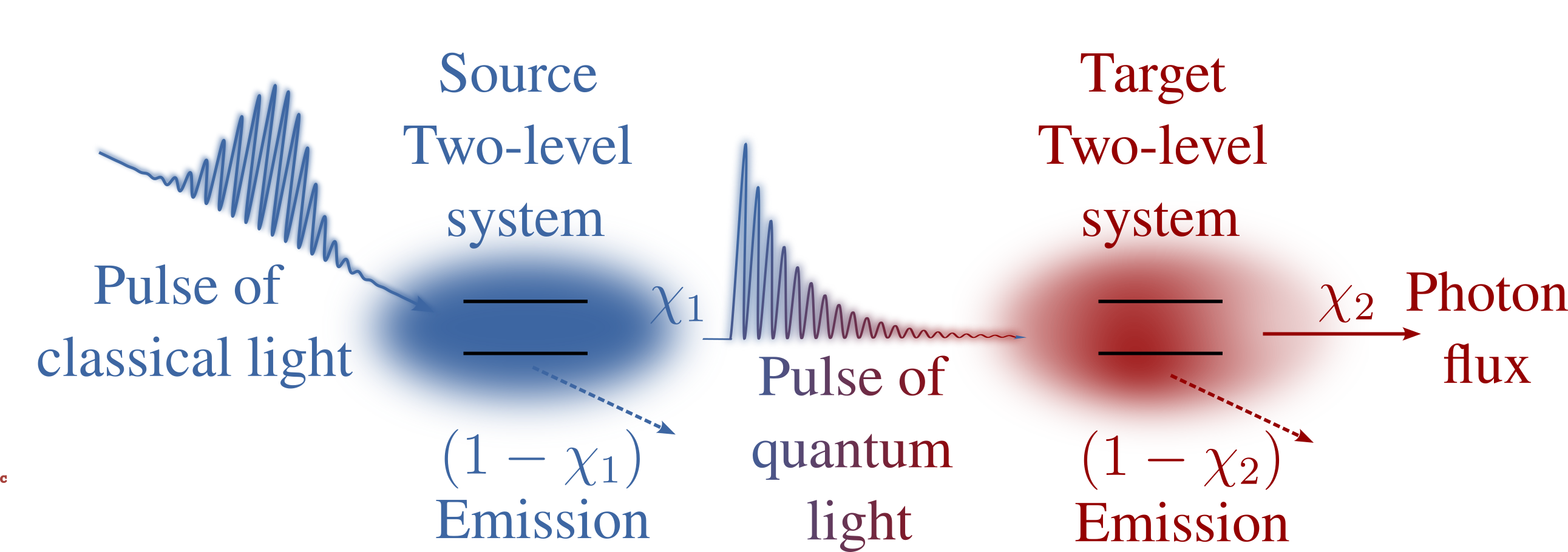}
  \end{center} 
  \caption{\textbf{Sketch of the cascaded excitation.} A classical
    pulse of light (blue line) is used to excite a two-level system,
    here depicted as a glowing halo around a pair energy levels. A
    fraction~$\chi_1$ of the emission from the source 2LS becomes
    the source of quantum excitation of the target 2LS (line with
    blue-to-red gradient coloring). After the interaction between
    the target 2LS and the quantum excitation we can collect two
    fields: namely, a fraction~$\chi_2$ corresponding to the photon
    flux of the entire system, and the remaining
    fraction~$(1-\chi_2)$ associated to the individual emission from
    the target 2LS.}
  \label{fig:WedNov6114531CET2024}
\end{figure}

From the point of view of the source of quantum light, the most
fundamental system that one can consider is Resonance
Fluorescence. Described as two energy levels that become dressed as
they interact with a strong coherent field, it yields
the celebrated \emph{Mollow triplet} spectrum~\cite{mollow1969}. Such a
system has a rich underlying structure~\cite{lopezcarreno2017},
whereby the photons emitted from various frequency regions display a
large gamut of correlations~\cite{gonzalez-tudela2013,
  peiris2015}. Using this variety of photon correlations to excite
fermionic~\cite{lopezcarreno2016a} and bosonic
fields~\cite{lopezcarreno2016} strengthens their quantum features
(for example, leading to better antibunching in the final emission)
and lets them access quantum states that are unreachable when they
are illuminated with classical light. Notably, the Mollow triplet
can be used to perform an unprecedented quantum spectroscopy
technique, whereby the internal structure of the target of the
excitation is probed with only two photons~\cite{lopezcarreno2015}.

A natural advancement in the analysis of Resonance Fluorescence was
to upgrade the continuous-wave driving, and consider pulsed
excitation instead. Under such a regime, the emission spectrum
consists of multiple peaks~\cite{rodgers1991, rzazewski1984,
  buffa1988, lewenstein1986, he2015, florjanczyk1985}, whose
linewidths and intensities depend on the strength and duration of
the pulse~\cite{moelbjerg2012, gustin2018a}. The appearance of the
multiple-peaked structure has recently been observed in
semiconductor~\cite{boos2024} and superconducting quantum
dots~\cite{ruan2024}, as well as in cavity-QED
systems~\cite{liu2024}. Importantly, the spectral correlations of
the emitted photons are still present under this regime of
excitation~\cite{konthasinghe2015, bermudezfeijoo2025}.

In this Article, we go beyond the analysis of pulsed Resonance
Fluorescence, and instead exploit its emission as a source of
quantum light. Using a two-level system (2LS) as the optical target
of the excitation, we present here a direct comparison between
classical and quantum pulsed excitation, and showcase the benefits
of the latter for the implementation of quantum technologies. The
choice of source and target considered in this Article allows us to
provide an in-depth analysis of a recent experimental observation of
the excitation of a quantum dot with pulse of quantum
light~\cite{hansen2024}. Thus, the rest of the manuscript is
organised as follows. Section~\ref{sec:MonMay5120400CEST2025}
provides the theoretical tools used to described the excitation of
the source 2LS with a laser pulse
(cf. \textbf{Figure~\ref{fig:WedNov6114531CET2024}}), and the subsequent
cascade of quantum light directed towards the target 2LS. Next, in
sections~\ref{sec:FriApr4114237CEST2025}---\ref{sec:FriApr4114323CEST2025}
we compare the single-particle observable of the emission,
considering the dynamic Rabi oscillations, the occupation and the
time-integrated emission spectrum. Later, in
section~\ref{sec:FriApr4114401CEST2025} we discuss the two-particle
observables. There, we present evidence of stimulated emission at
the single-particle level and show that quantum excitation yields an
enhancement in the antibunching and the indistinguishability of the
emitted photons. Finally, in section~\ref{sec:FriApr4114458CEST2025}
we conclude and present the perspectives of our work.

\section{Theoretical description}
\label{sec:MonMay5120400CEST2025}

The first stage of the excitation, used to power the source of
quantum light, is done with laser pulse and it is
modelled through the Hamiltonian (we take~$\hbar=1$ along the
manuscript)
\begin{equation}
  \label{eq:Wed8Nov2023135425CET}
  H_\sigma(t) = (\omega_\sigma - \omega_\mathrm{L}) \ud{\sigma} \sigma
  + \frac{\Omega (t)}{2} (\sigma + \ud{\sigma})\,,
\end{equation}
where~$\sigma$ is the annihilation operator associated to the 2LS
(and which follows the pseudo-spin algebra). Here, we have also
defined the natural frequency of the 2LS~$\omega_\sigma$, the
frequency of the pulse laser~$\omega_\mathrm{L}$, and the
time-dependent intensity of the pulse~$\Omega(t)$. The dissipative
character of the system is taken into account by upgrading the
description to a master equation
\begin{equation}
  \label{eq:Wed8Nov2023140117CET}
  \partial_t \rho = i \left[\rho, H_\sigma(t) \right] +
  \frac{\gamma_\sigma}{2} \mathcal{L}_\sigma (\rho)\,,
\end{equation}
where~$H_\sigma(t)$ is the Hamiltonian introduced in
\textbf{Equation~(\ref{eq:Wed8Nov2023135425CET})},
$\mathcal{L}_c (\rho) = (2 c\rho \ud{c} - \ud{c}c\rho - \rho
\ud{c}c)$, and $\gamma_\sigma$ is the decay rate of the
2LS. Commonly, the laser pulse has a Gaussian profile, given by
\begin{equation}
  \label{eq:Wed8Nov2023144540CET}
  \Omega(t) =  \frac{A}{\sqrt{2 \pi \nu^2}} \exp
  \left[-\frac{(t-t_0)^2}{2 \nu^2} \right]\,,
\end{equation}
which corresponds to a pulse with integrated area~$A$, centered at
time~$t_0$ and with variance~$\nu^2$. The latter is related to the full
width at half maximum (FWHM) through the
relation~$\mathcal{W} = 2\sqrt{2\log 2}\nu$; and along this
manuscript we will refer to $\mathcal{W}$ as the \emph{length of the
pulse}.

\begin{figure*}[b]
  \includegraphics[width=\linewidth]{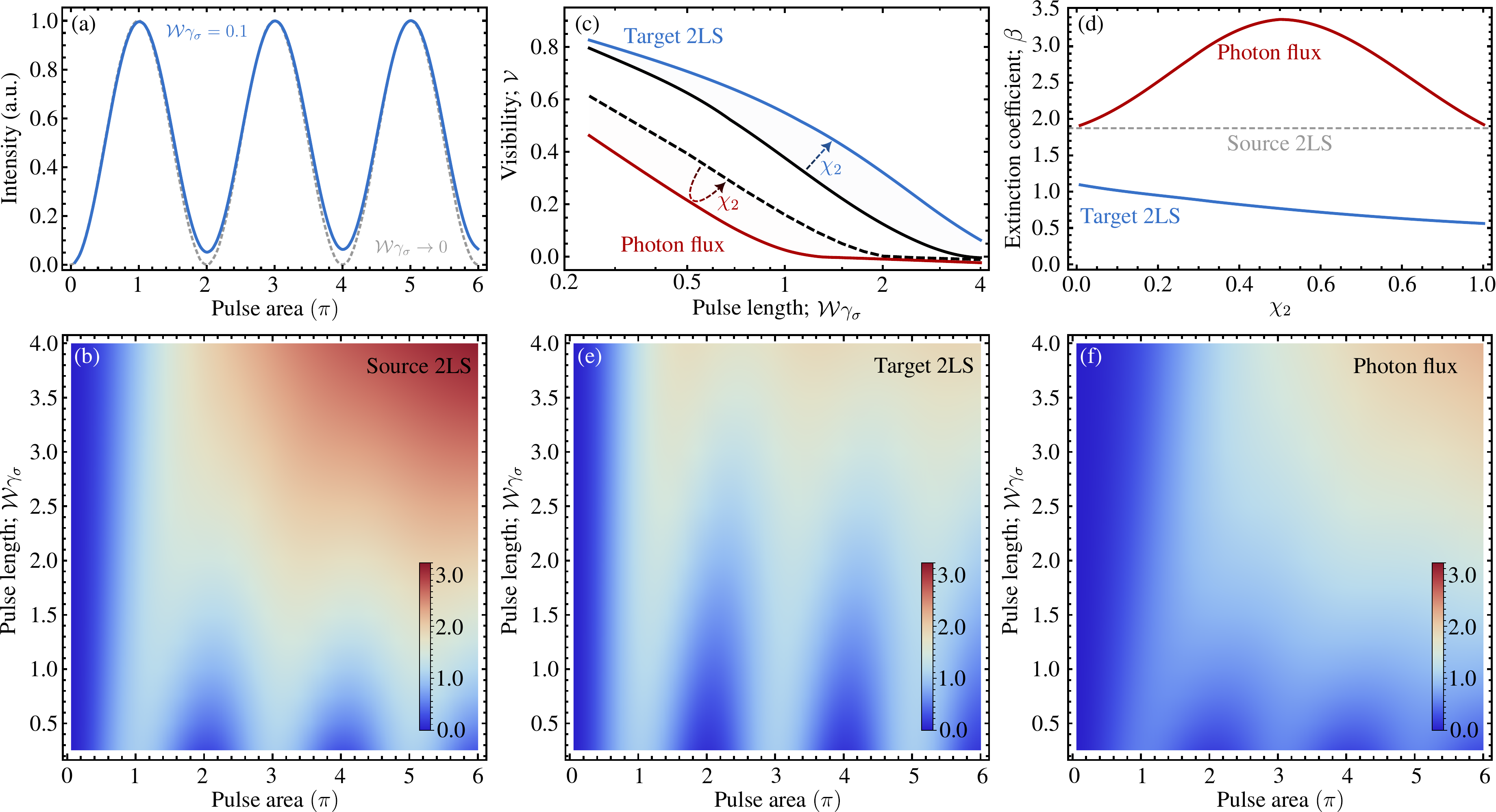}
  \caption{\textbf{Pulsed Rabi oscillations.} (a)~Intensity of the
    emission from a 2LS pulsed with laser with vanishing (dashed
    line) and finite but short length (solid line). (b)~Intensity of
    the emission of the source 2LS as a function of both the area
    and the length of the pulse, showing that the Rabi oscillations
    are quickly suppressed as the length of the pulse is
    increased. (c)~Visibility of the Rabi oscillations, as
    quantified by the normalized difference between the maximum
    near~$A=\pi$ and the minimum near~$A=2\pi$, for the source 2LS
    (dashed), the target 2LS (ranging between the upper two lines,
    from solid black to solid blue, covering the entire area shaded
    in blue) and the photon flux (ranging between the lower two
    lines, from dashed black to solid red, convering the entire area
    shaded in red). For the latter two, the gradient indicates the
    variation of the parameter~$\chi_2$. (d)~Extinction coefficients
    for the visibility lines showed in panel~(c), for the source 2LS
    (dashed), the target 2LS (blue) and photon flux (red). (e,
    f)~Same as in panel (b) but for the target 2LS and the photon
    flux, respectively; calculated for~$\chi_2=1/2$. The panels in
    the bottom row share the same color encoding, and therefore can
    be compared directly.}
  \label{fig:FriOct11160737CEST2024}
\end{figure*}

In the second stage of the excitation, the emission from the
classically excited 2LS is directed towards the second 2LS which, as
a consequence, is driven with a quantum field. (The details of the
experimental setup to realise such a quantum excitation can be found
in the Supplemental Material of Ref.~\cite{hansen2024}. Note that in
that particular experiment the quantum emitter was embedded in a
microcavity. However, the role of the latter was purely technical;
namely, to enhance the efficiency of excitation of the emitter, and
did not have any other effect in the dynamics of the quantum
emitter. Therefore, our description can safely omit the microcavity
and still reproduce the experimental observations.)  From a
theoretical point of view, such an excitation is a perfect example
of the dynamics that inspired the theory of cascaded
systems~\cite{gardiner1993, carmichael1993}. Therefore, the master
equation that describes the full system, including the quantum
excitation, and which reproduces the experimental results of
Ref.~\cite{hansen2024}, becomes
\begin{multline}
  \label{eq:Wed8Nov2023141814CET}
\partial_t \rho = i
\left[\rho, H_\sigma(t) + H_\xi \right] + \frac{\gamma_\sigma}{2}
\mathcal{L}_\sigma (\rho) +
\frac{\gamma_\xi}{2} \mathcal{L}_\xi (\rho) - {}\\
{} - \sqrt{\chi_1 \chi_2 \gamma_\sigma \gamma_\xi} \left \lbrace
    \left[ \ud{\xi}, \sigma \rho\right] + \left[\rho \ud{\sigma},
      \xi \right] \right \rbrace\,.
\end{multline}
Here, $\xi$ is the annihilation operator of the target 2LS, which
decays with rate $\gamma_\xi$. The
Hamiltonian~$H_\xi = (\omega_\xi - \omega_\mathrm{L}) \ud{\xi}\xi$
describes the free energy of the target 2LS, which also showcases
the detuning between the natural frequency of the target 2LS
$\omega_\xi$ and the laser driving the source 2LS. The excitation
with quantum light is modelled in
\textbf{Equation~(\ref{eq:Wed8Nov2023141814CET})} through the term
in the second line. It provides an unidirectional coupling between
the source and target, in such a way that the dynamics of the former
is not affected by the presence of the latter. Namely, there is no
back-action from the target to the source of quantum light. Here it
is important to note that when the emitters are far enough from each
other, so that their dipole interaction can be omitted, but closer
than the coherence length of their emission, one could enter regimes
of \emph{delayed feedback}~\cite{grimsmo2015, whalen2017}. In such
cases, the dynamics of the system becomes
non-Markovian~\cite{breuer2016, devega2017, sinha2020}, which has
been investigated in the context, e.g., of the variation of the rate
of spontaneous emission of individual atoms~\cite{cook1987,
  dorner2002, tufarelli2013, carmele2013a, guimond2016a}, bound
states in the continuum~\cite{hsu2016, fong2017, facchi2019} and
entanglement generation with emitters coupled to
waveguides~\cite{gonzalez-ballestero2013, pichler2016,
  qiu2024}. While these phenomena are interesting on their own
merits, in this manuscript we remain in the regime where the target
2LS does not give any feedback to the source 2LS.

Turning back to our master equation, the parameters $\chi_1$ and
$\chi_2$ that appear in the coefficient of this coupling are related
to the degree to which each of the 2LS is able to couple to the
other one. This is more easily seen when the master
equation~(\ref{eq:Wed8Nov2023141814CET}) is rearranged in the
Lindblad form, and we interpret the result in terms of quantum jump
operators~\cite{dalibard1992}. (See, e.g., the Supplemental Material
of Ref.~\cite{lopezcarreno2024} for the details of the general
transformation, and the Supporting Information for a detailed
derivation of the equations presented here.) Then, we find that the
emission from the 2LSs is ruled by the operators
\begin{equation}
  \label{eq:FriApr11124401CEST2025}
  J_\sigma = \sqrt{(1-\chi_1)\gamma_\sigma}\sigma \quad \quad \mathrm{and}
  \quad \quad J_\xi = \sqrt{(1-\chi_2)\gamma_\xi}\xi\,.
\end{equation}
In practice, this means that, e.g., the intensity of the emission
from the source 2LS is decreased by a factor~$(1-\chi_1)$, because
the rest of the energy is used to excite the other 2LS. Finally,
there is a third Lindblad term that mixes the operators from the
two 2LS. Namely, $J_\phi$ defined as 
\begin{equation}
  \label{eq:Wed8Nov2023143727CET}
  J_\phi = \sqrt{\chi_1 \gamma_\sigma} \sigma + \sqrt{\chi_2
    \gamma_\xi} \xi\,,
\end{equation}
which, is commonly referred to as the \emph{photon
  flux}~\cite{kiilerich2019, kiilerich2020} and, as we shall see in
the following sections, induces a quantum interference that provides
a mechanism to stimulated emission of the target 2LS.

From the decomposition above we can clearly see the interplay
between the strength of the cascaded coupling
$\sqrt{\chi_1 \chi_2 \gamma_\sigma \gamma_\xi}$ and the jump
operators of the system. Namely, neither $\chi_{1}$ nor $\chi_2$ can
be zero, because it would mean that the two 2LS are not coupled. On
the other hand, if they are set to 1, then the jump operators in
Eq.~(\ref{eq:FriApr11124401CEST2025}) are zero. In principle, this
is not a problem. However, it implies that one cannot observe the
emission from the 2LSs ``on their own'', and instead one can only
see the collective emission through the photon flux. Given that we
can collect the information from the source 2LS independently (that
is, without the cascaded coupling, because the presence, or absence,
of the optical target does not affect the dynamics of the source
2LS), and that for the target 2LS using $0 \leq \chi_1 <1$ is
equivalent to using a weaker source, we will consider the
case~$\chi_1=1$. In our particular case, letting~$\chi_1 = 1$ means
that the laser light is completely removed from the observables of
the 2LSs. Therefore, unlike the experimental case, our theoretical
description does not require any further technique to prevent
contamination from the source of classical excitation.  On the other
side, for the target 2LS we want to keep the ability to observe the
field on its own, as well as on the photon flux of the system. Thus,
as a compromise between the two dissipative channels, in the main
text we will set~$\chi_2=1/2$. However, in the Supporting
Information we show how our results change for other values
of~$\chi_2$. In practice, $\chi_1$ is the fraction of the emission
of the source 2LS that does not carry scattering or photons from the
driving laser. Namely, the fraction of emission that is originating
only from the luminescence of the 2LS. Similarly, the emission from
the target 2LS can also be split into a fraction~$(1-\chi_2)$
containing only its luminescence, and the remaining
fraction~$\chi_2$ that consists of the photon flux of the entire
system; namely, part of the luminescence of the target 2LS together
with scattered photons from the source of excitation. In the
following, therefore, we will show the behaviour of the luminescence
of a 2LS driven with classical light (labelled as \emph{Source
  2LS}), the luminescence of a 2LS driven with quantum light
(labelled as \emph{Target 2LS}), and the field observed
experimentally in Ref.~\cite{hansen2024}, namely, the superposition
of the luminescence of the target 2LS and the (quantum) excitation
field (labelled as \emph{Photon flux}).

\begin{figure*}[t]
  \includegraphics[width=\linewidth]{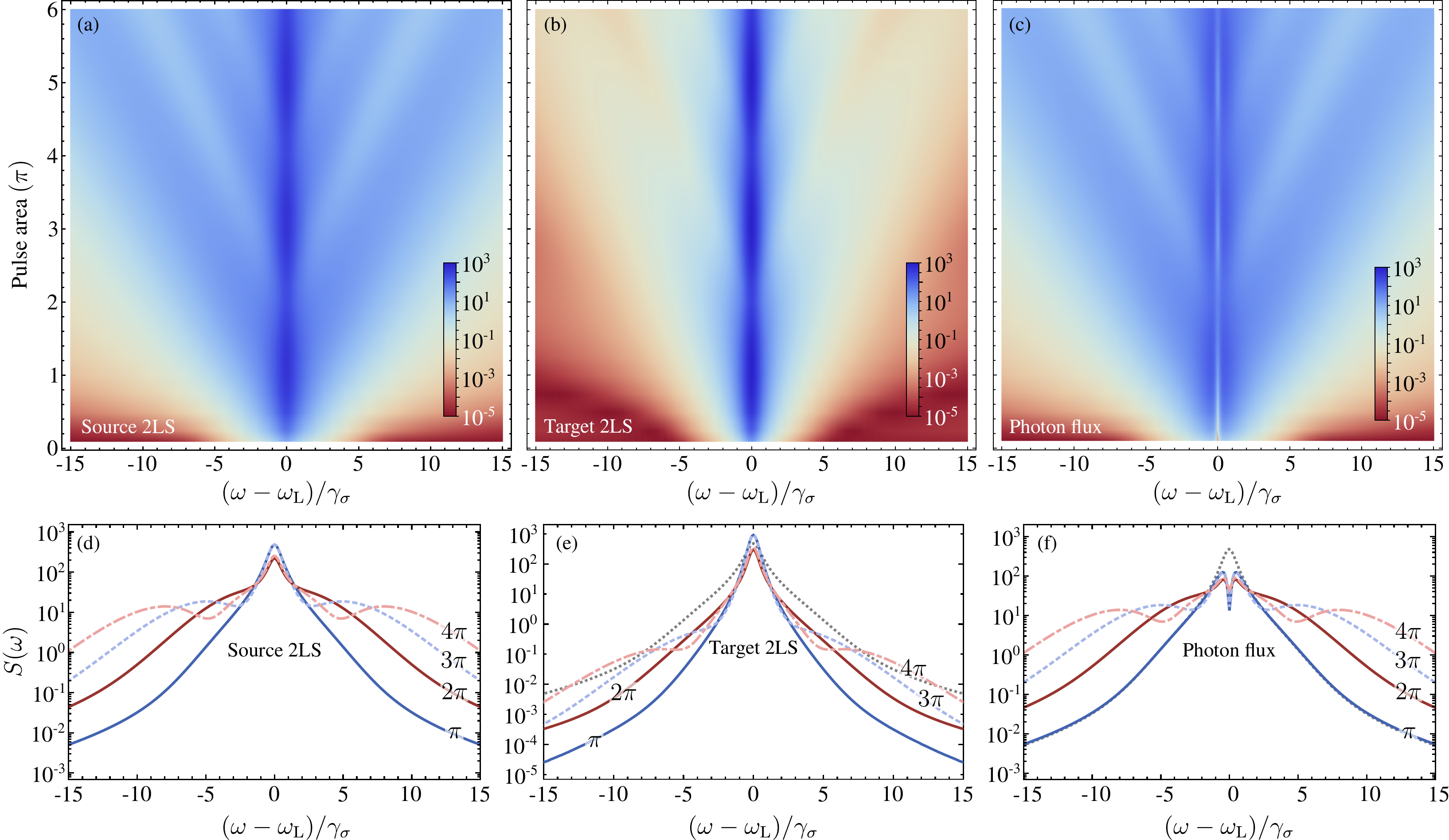}
  \caption{\textbf{Emission spectra of dynamic Resonance
      Fluorescence.}  (a--c)~Photoluminescence as a function of the
    area of the pulse, for the source 2LS, target 2LS and the photon
    flux, respectively. The variation in the intensity of the
    emission as the pulse area increases is an echo of the Rabi
    oscillations undergone by the 2LS.  (d--f)~Cuts of the spectra
    for $\pi$- (solid blue), $2\pi$- (solid red), $3\pi$- (dashed
    light blue) and $4\pi$-pulses (dot-dashed light red). In panels
    (e) and (f) the dotted gray line corresponds to the PL of the
    source 2LS for a $\pi$-pulse, showing the line narrowing of the
    target 2LS. The figures are obtained for pulses at resonance to
    the 2LSs, and setting~$\mathcal{W}\gamma_\sigma = 1$
    and~$\chi_2=1/2$. The three upper panels can be compared
    directly, as they share the same color scaling.}
  \label{fig:FriOct11154401CEST2024}
\end{figure*}

\section{Pulsed Rabi Oscillations}
\label{sec:FriApr4114237CEST2025}

The response of the classically excited 2LS as a function of the
area of the laser pulse is well known~\cite{allen1987,
  steck2024}. Namely, for short-enough pulses, once the 2LS has
interacted with the entirety of the pulse, the probability of
finding the 2LS on its excited state is given by
\begin{equation}
    \label{eq:FriOct4163716CEST2024}
    P_e = \sin^2(A/2)\,,
\end{equation} 
where~$A$ is the integrated area of the pulse. This is known as the
area theorem.

Thus, one expects Rabi oscillations with unit visibility as a
function of the area of the pulse, as shown in the dashed line in
\textbf{Figure~\ref{fig:FriOct11160737CEST2024}}(a). However,
although \textbf{Equation~(\ref{eq:FriOct4163716CEST2024})} is a
good approximation for very short pulses, even
for~$\mathcal{W}\gamma_\sigma =0.1$ we find a deviation from the
perfect oscillatory behaviour~\cite{fischer2016a, fischer2017a,
  masters2023}, as shown as a solid line
Figure~\ref{fig:FriOct11160737CEST2024}(a). In fact, as the length
of the pulse is increased, the oscillations become increasingly
damped, up to the point in which the intensity of the light
increases monotonically as a function of the pulse area. Such a
behaviour is illustrated in
Figure~\ref{fig:FriOct11160737CEST2024}(b), where we show the
integrated intensity as a function of the pulse area. Namely,
  the panels in the lower row display
\begin{equation}
  \label{eq:TueSep2152332CEST2025}
  I_c = \int_0^{\infty}\mean{\ud{c}c} (t)\, dt\,,
\end{equation}
for $c = \lbrace \sigma, \xi, J_\phi\rbrace$, respectively. The
upper limit of integration is taken so that the entire response of
the field is captured. There we can also see that, although for
short pulses the position of the minima and maxima of the
oscillations are given by odd- and even-$\pi$ pulses, respectively,
as the length of the pulses increases, these positions drift
away. Therefore, the length of the pulse, together with the decay of
the emitter and its coupling to phonons~\cite{steck2024}, is yet
another cause of the shift of the maxima of the Rabi oscillations.

Further, we can quantify the visibility the oscillations through the
ratio
\begin{equation}
  \label{eq:TueOct15123615CEST2024}
  \mathcal{V} = \frac{M_{\pi}-m_{2\pi}}{M_{\pi}+m_{2\pi}}\,,
\end{equation}
where~$M_{\pi}$ is the maximum of the intensity around $\pi$-pulses,
and $m_{2\pi}$ is the minimum of the intensity
around~$2\pi$-pulses. For the 2LS driven classically, the visibility
of the oscillations is shown as a dashed line in
Figure~\ref{fig:FriOct11160737CEST2024}(c), which displays and
exponential decay as a function of the pulse length; namely, the
visibility of the oscillations can be fitted as
\begin{equation}
  \label{eq:TueOct15124452CEST2024}
 \mathcal{V} \propto \exp(-\beta \mathcal{W}\gamma_\sigma)\,,
\end{equation}
where~$\beta$ is the extinction coefficient, which for the source
2LS we fitted to~$\beta = 1.87$. Further, the dashed black line in
Figure~\ref{fig:FriOct11160737CEST2024}(c) shows that the oscillations
are completely extinguished for pulses of
length~$\mathcal{W}\gamma_\sigma \geq 2$.

We now turn to the Rabi oscillations of the target 2LS and the
photon flux, described with the operators~$\xi$ and~$J_\phi$,
introduced in Equation~(\ref{eq:Wed8Nov2023141814CET}) and
\textbf{Equation~(\ref{eq:Wed8Nov2023143727CET})}, respectively. The
oscillations of the target 2LS are shown in
Figure~\ref{fig:FriOct11160737CEST2024}(e), where we find that they
are much more robust to the length of the classical pulse, remaining
visible for pulses of length of up to four lifetimes of the 2LS, as
shown in the blue shaded region of
Figure~\ref{fig:FriOct11160737CEST2024}(c). There, the upper two
lines show the visibility for~$\chi_2 \rightarrow 0$ (black)
and~$\chi_2\rightarrow 1$ (blue), and the color gradient indicates
the variation of the parameter of the cascaded
coupling~$\chi_2$. Notably, as shown in
Figure~\ref{fig:FriOct11160737CEST2024}(c) and
\ref{fig:FriOct11160737CEST2024}(d), the Rabi oscillations of the
target 2LS are always more visible and decay more slowly than those
of the source 2LS.
Conversely, for the photon flux, whose oscillations are shown in
Figure~\ref{fig:FriOct11160737CEST2024}(f), we find that their
oscillations are, at best, as visible as the oscillations of the
source 2LS. The lower two lines in
Figure~\ref{fig:FriOct11160737CEST2024}(c) show the visibility
for~$\chi_2 \rightarrow 0$ (dashed black, which corresponds to the
visibility of the source 2LS) and~$\chi_2\rightarrow
1/2$~(red). Note that for the photon flux, as~$\chi_2>1/2$, the
visibility of the oscillations increases again. Finally, we can also
fit the visibility lines with
\textbf{Equation~(\ref{eq:TueOct15124452CEST2024})}, thus finding
the extinction coefficient as a function of the~$\chi_2$ parameter;
showing that the visibility in the oscillations for the target 2LS
and the photon flux is optimized reaching the
limit~$\chi_2 \rightarrow 1$.

\section{Time-integrated emission spectra}
\label{sec:FriApr4114214CEST2025} 
Beyond the intensity of the light emitted by the 2LS, we now
consider the spectral structure of the emission. The time-integrated
emission spectra of the various objects in our quantum system are
obtained through the calculations of the first-order correlation
function~$G_c^{(1)}(t,\tau) = \mean{\ud{c}(t+\tau)c(t)}$, where
$c=\lbrace \sigma, \xi, J_\phi\rbrace$, and then computing the
Fourier transform with respect to the delays~$\tau$ and integrating
over the evolution times~$t$; namely
\begin{equation}
  \label{eq:ThuApr10101700CEST2025}
  S(\omega) = \mathrm{Re} \left [ \int_{-\infty}^\infty dt
    \int_{-\infty}^\infty d\tau\,   G_c^{(1)}(t,\tau) e^{-i\omega
      \tau}\right]\,. 
\end{equation} 
Although the time-integrated emission spectrum of a 2LS driven with
pulses of classical light has been already worked out
theoretically~\cite{florjanczyk1985, moelbjerg2012, gustin2018a} and
observed~\cite{fischer2016a, fischer2017a, masters2023}, here we
reproduce it in \textbf{Figure~\ref{fig:FriOct11154401CEST2024}}(a)
so that we can compare it with the spectra of the target 2LS and the
photon flux. In particular, we observe that the central frequency
remains dominant for all pulse areas, along with the lateral peaks
arising for $2n\pi$ pulses as a consequence of the dynamic state
dressing~\cite{moelbjerg2012, gustin2018a,
  boos2024}. Figure~\ref{fig:FriOct11154401CEST2024}(d) shows cuts
of the density plot for pulse areas $A=\pi$ (solid blue), $2\pi$
(solid red), $3\pi$ (dashed light blue) and $4\pi$ (dot-dashed light
red), evidencing the birth, growth and separation of the side peaks
as the intensity of the laser increases.

The emission spectra of the target 2LS (which is driven with pulses
of quantum rather than classical light) is shown in
Figure~\ref{fig:FriOct11154401CEST2024}(b), where it appears to be
given by a single line of oscillating intensity. However, a closer
look in logarithmic scale, shown in
Figure~\ref{fig:FriOct11154401CEST2024}(e), reveals that the dynamic
dressing of the states of the source 2LS is also imprinted onto the
spectrum of the target 2LS. However, in this case, the side bands
are much less prominent. Notably, we find that the emission line of
the source 2LS, shown as a dotted grey line, is broader that the
lines of the target 2LS.  As previously observed with quantum
excitation in the continuous regime~\cite{lopezcarreno2018a,
  lopezcarreno2019}, the target 2LS effectively behaves as a filter
(of linewidth~$\gamma_\xi$) of the emission of the source 2LS, thus
trimming the fat tails of the Lorentzian profile of the latter. As a
consequence, the emission line of the target 2LS is narrower than
its natural linewidth~$\gamma_\xi$ and it profile is given by a
Student-$t$ distribution of second
order~\cite{lopezcarreno2018a}. Therefore, we find that the pulsed
quantum cascaded excitation also induces an spectral line narrowing
in the target 2LS.

Finally, the emission spectrum of the photon flux, shown in
Figure~\ref{fig:FriOct11154401CEST2024}(c), closely resembles its
counterpart for the source 2LS. However, we see that exactly at the
resonant frequency of the emitters there is a dip in the intensity
of the emission: these are the photons that, instead of being
observed through the photon flux, are observed directly from the
target 2LS. However, the side peaks appear more prominently, because
for the photon flux the maximum of the central peak is reduced.  The
cuts of the density plot are shown in panel~(f) and show that the
photon flux does not have any line narrowing (cf. the dotted grey
line, corresponding to the PL of the source 2LS for $\pi$-pulse
excitation), and that the absolute intensity of the side peaks is
maintained, which remain as intense as in the figures for the source
2LS.

\section{Time-dependent occupation}
\label{sec:FriApr4114323CEST2025}

The population of the source 2LS can be described qualitatively as
follows: odd-$\pi$ pulses leave the 2LS in its excited state,
followed by the exponential decay associated to the spontaneous
emission from the 2LS. Conversely, even-$\pi$ pulses (almost) leave
the 2LS in its ground state. However, the latter statement (without
the ``almost'') is only true in the limit of vanishingly short
pulses. Beyond that limit, the 2LS is left in a superposition of its
excited and ground states. Therefore, for even-$\pi$ pulses of
finite length, one also observes an exponential decay in the
occupation of the 2LS after it has interacted with the whole pulse.

\begin{figure}
  \begin{center}
    \includegraphics[width= \linewidth]{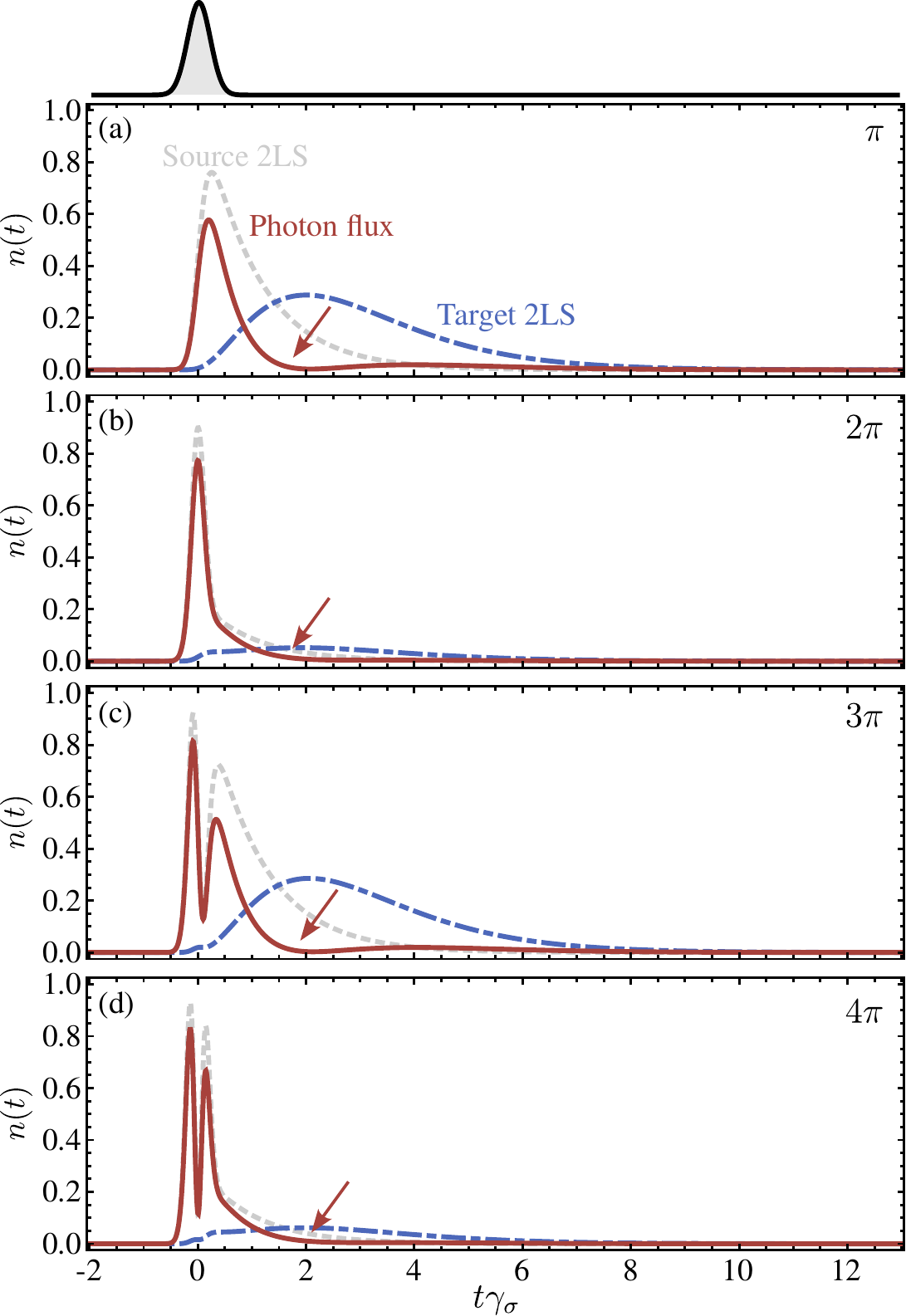}
  \end{center}
  \caption{\textbf{Time-dependent occupation of the pulsed
      emitters.} Comparison between the averaged time-dependent
    population of the source 2LS (dashed gray), the photon flux
    (solid red) and the target 2LS (dot-dashed blue), for various
    pulse areas (noted on the top right corner of each panel). All
    the panels are obtained with a pulse of length
    $\mathcal{W}\gamma_\sigma = 1/2$ (which for reference is
    sketeched at the top of the figure) and letting~$\chi_2 =
    1/2$. The underlying quantum interference taking place in the
    field of the photon flux is particularly noticeable in
    panels~(a) and~(c), where a dip in the occupation can be
    distinguished at $t\gamma_\sigma \sim 2$. In all the panels,
    the origin of time is set by the position of the maximum of the
    classical pulse.}
  \label{fig:ThuApr10115912CEST2025}
\end{figure}

The population of the source 2LS (dashed grey), the target 2LS
(dot-dashed blue) and photon flux (solid red) are shown in
\textbf{Figure~\ref{fig:ThuApr10115912CEST2025}} for four pulse
areas. On top of the figure there is a sketch of the Gaussian laser
pulse exciting the source 2LS. In general, as we found in the
previous section, the figures for the photon flux follow closely the
behaviour of the source 2LS. However, there are notable
differences. Firstly, the curves for the photon flux are less
intense, because part of the energy from the source 2LS goes to the
target 2LS. Secondly, the populations for the photon flux display a
small, broad peak beyond the \emph{main} features of the
populations. Namely, after the pulse is absorbed, the population
does not decrease monotonously. Instead, we see a dip in the
occupation (indicated by the arrows), followed by a second peak a
few lifetimes later. Although this feature is more prominent for
odd-$\pi$ pulses, it is observed for all pulse areas, even for long
pulses (cf. the Supporting Information for an analysis of the effect
of the length of the pulse and the coupling parameter~$\chi_2$ on
the occupations). Formally, the appearance of the delayed peak is a
consequence of the interference between the emission from the source
and from the target 2LSs, which becomes evident when computing the
population of the photon flux,
namely~$\Tr \{ \rho(t) J_\phi^\dagger J_\phi \}$, with~$J_\phi$ the
operator in Equation~(\ref{eq:Wed8Nov2023143727CET}). On the other
hand, the population of the target 2LS has a less intense profile,
which grows and decays smoothly, and extends further in time. Here
we no longer see the rapid oscillations induced by the laser pulse,
which allows the photons be, on average, further apart from each
other than in the field of the source 2LS or of the photon flux. In
the next section, we will explore in detail what this implies in
terms of the photon antibunching and indistinguishability of the
emission.

\section{Second-order correlations}
\label{sec:FriApr4114401CEST2025}

\subsection{Enhancement of the antibunching}

\begin{figure}[b]
  \begin{center}
    \includegraphics[width=\linewidth]{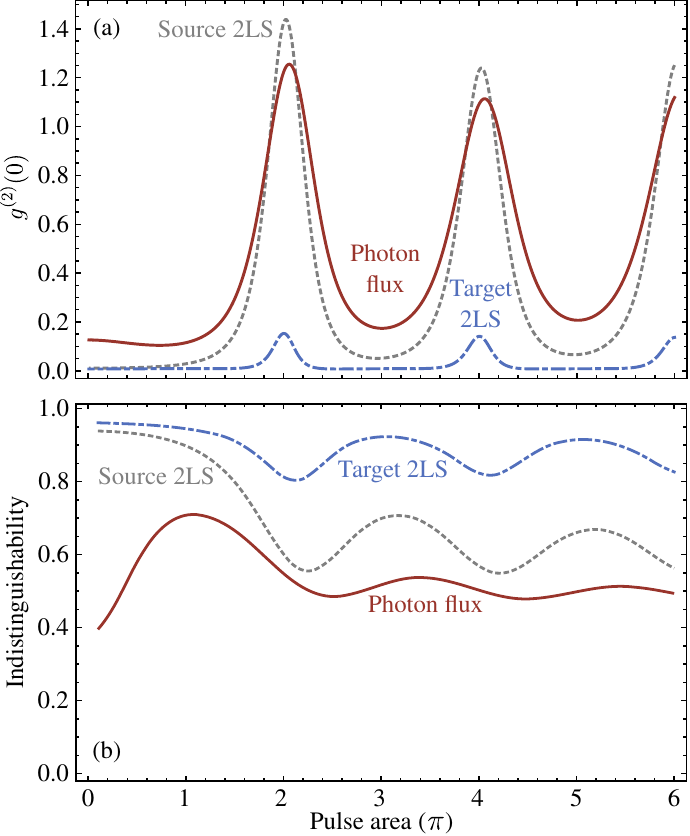}
  \end{center}
  \caption{\textbf{Second-order correlation functions of the emitted
      photons.} (a)~Zero-delay~$g^{(2)}(0)$ and (b)~Photon
    indistinguishability, as captured by the visibility of the dip
    in a HOM interference. Here, we show the information for the
    source 2LS (dotted gray), the target 2LS (dot-dashed blue) and
    the photon flux (solid red). In both cases, the figures are
    obtained with a pulse of $\mathcal{W}\gamma_\sigma = 0.25$
    and letting~$\chi_2 = 1/2$.}
  \label{fig:TueApr15124341CEST2025}
\end{figure}

We now turn to two-photon observables and delve into the temporal
structure of the light. Specifically, we consider the second-order
coherence function~$g^{(2)}(0)$ of the
emission. \textbf{Figure~\ref{fig:TueApr15124341CEST2025}}(a) shows
the correlation function of the source 2LS (dashed grey), the photon
flux (solid red) and the target 2LS (dot-dashed blue) as a function
of the area of the pulse. 

\begin{figure*}[b]
  \begin{center}
    \includegraphics[width=0.85\linewidth]{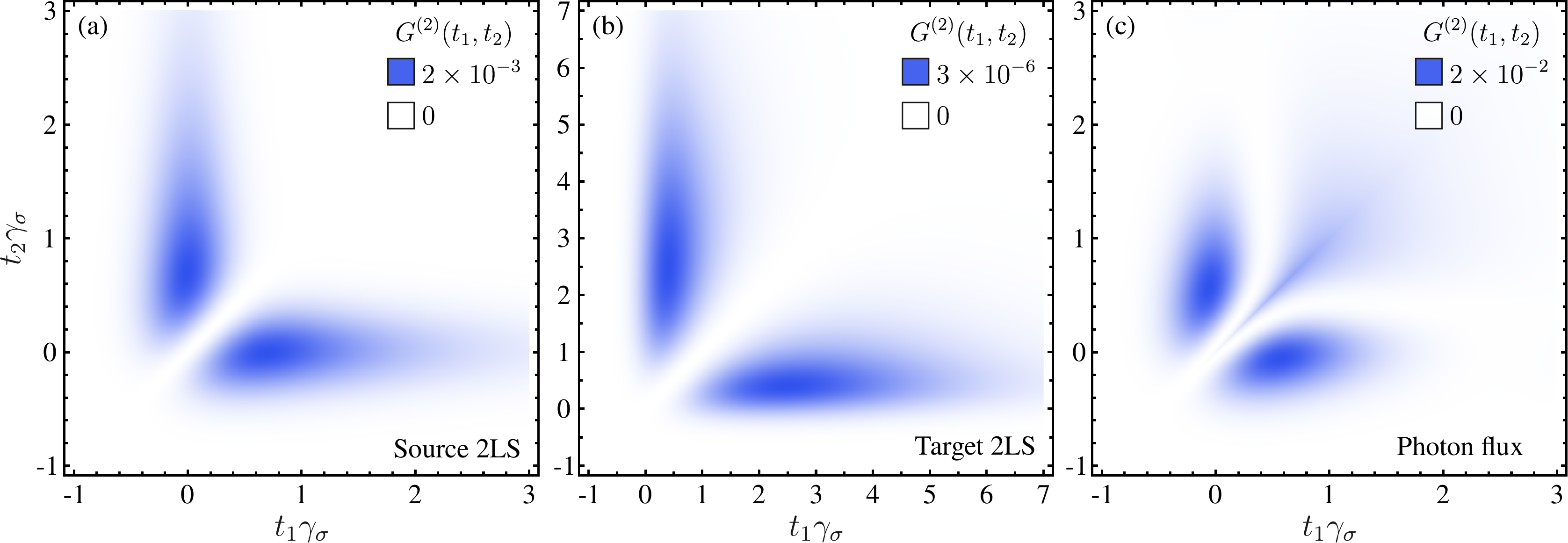}
  \end{center}
  \caption{\textbf{Time-delayed second-order coherence
      function~$G^{(2)}(t_1, t_2)$.} (a)~Correlation function for
    the source 2LS, displaying correlations between photons emitted
    in the close vicinity of the maximum of the excitation
    pulse. (b)~The target 2LS inherits the shape of its correlation
    from the source 2LS, but although they have the same decay
    rate~$\gamma_\sigma$, the correlations of the target 2LS spread
    further in time and are attenuated by over three orders of
    magnitude. (c)~The photon flux also has the lobes around the
    maximum of the excitation pulse, but displays correlations on
    the diagonal of the figure, which correspond to photons emitted
    simultaneously ($t_1=t_2$) and beyond the lifetime of the
    2LS. The figure corresponds to excitation with a $\pi$-pulse of
    length~$\mathcal{W}\gamma_\sigma =1$. In all the panels,
    the origin of time is set by the position of the maximum of the
    classical pulse.}
  \label{fig:TueNov5104935CET2024}
\end{figure*}

The figures show the ratio of the correlations between photons
emitted from a single pulse and the correlations between photons
emitted from different pulses. Namely, the normalization commonly
done in experiments with pulsed excitation. For the source 2LS,
shown as gray dashed line in
Fig.~\ref{fig:TueApr15124341CEST2025}(a), we find a well known
result: the correlations are bunched for even-$\pi$ pulses and
antibunched for odd-$\pi$ pulses. This is because for even-$\pi$
pulses, the probability to measure two photons $P_2$ is larger than
the probability to measure only one photon
$P_1$~\cite{fischer2017a}, while the overall intensity of the
emission is at its minimum
[cf. Fig.~\ref{fig:FriOct11160737CEST2024}(a)]. The combination of
these two facts yields bunching among the emitted photons. For
odd-$\pi$ pulses the reverse is true: $P_1 > P_2$, the intensity of
the emission is at its maximum, and as a consequence we observe
antibunching. In the intermediate pulse areas, the system undergoes
a smooth transition between these two limits. Considering now the
photon flux of the system, shown in solid red, we find a nice match
with the recent experimental observation~\cite{hansen2024},
confirming that the output field measured in the experiment was
actually the photon flux of the system, and not only the
luminescence from the target 2LS. Thus, the photons of the photon
flux are always less antibunched than those of the source 2LS,
except at a region in the vicinity of $2\pi$ (from our theoretical
calculation, we find that this is true in the vicinity of $2n\pi$),
where the relation is reversed. This is because the incoming
temporally separated two-photon components are partially changed to
one-photon components in the detected output field, leading to
lower~$g^{(2)}(0)$ values in the photon
flux~\cite{hansen2024}. However, at such pulse areas, the photons
are bunched, so there is no enhancement of the single-photon
emission. However, turning to the emission from the target 2LS, we
find that their photon are \emph{always more antibunched} than those
emitted by the source 2LS. This observation is in agreement with
previous theoretical predictions made on cascaded structures excited
with cw-excitation~\cite{lopezcarreno2016a}. Notably, even for
$2n\pi$ pulses, where the photons of both the source 2LS and the
photon flux are bunched, the emission from the target 2LS remains
antibunched. Thus, here we provide evidence that the quantum
cascaded excitation is a mechanism to improve the antibunching in
pulsed systems. (In the Supporting Information we show how these
correlations vary depending on the length of the classical pulse,
and that our results hold even with ultra-short pulses, used in
state-of-the-art setups to extract single photons
\cite{somaschi2016, tomm2021}.) Namely, as the pulses become
broader, the maxima of the correlations, which still occur for
pulses of (approximately) even-$\pi$ areas, remain antibunched,
i.e., below 1. Thus, even if the emission for, e.g., 2~$\pi$ pulses
is expected to be a bunched pair of photons, this is only true in
the limit of small pulse lengths. When the latter condition is not
fulfilled, the photons pairs are less antibunched than those emitted
from, e.g., $\pi$-pulses, but remain antibunched nonetheless. In the
limit of very long pulses, where the excitation profile becomes
essentially flat, we recover the results of cw-excitation; namely,
perfect antibunching in the source 2LS, target 2LS and photon
flux. Moreover, in the same way in which the maxima and minima of
the Rabi oscillations drift away from integer-$\pi$-pulses, as the
length of the pulses increases, the maxima of correlations also
drift away from the even-$\pi$-pulses; which is specially visible
for the case of the photon flux (cf.~the Supporting Information).

\subsection{Enhancement of the indistinguishability}

Once we have established that the quantum excitation yields an
enhancement in the antibunching of the emission, we address another
important characteristic of sources of single photons, namely their
indistinguishability. Such a property is measured with a
Hong-Ou-Mandel (HOM) interferometer~\cite{hong1987}, whereby the
visibility of the interference is directly related to the
indistinguishability of the input states~\cite{bylander2003,
  legero2003, trivedi2020}, even for imperfect sources of single
photons~\cite{ollivier2021}.
Figure~\ref{fig:TueApr15124341CEST2025}(b) shows the
indistinguishability of the photons emitted by the source 2LS
(dashed gray), the target 2LS (dot-dashed blue) and the photon flux
(solid red) as a function of the area of the pulse. Comparing the
values obtained for $\pi$-pulses, we find a very good agreement with
the recent experimental observations~\cite{hansen2024}: the source
2LS yields photons with a $\sim 91\%$ indistinguishability, while
for the photon flux this value decreases to $\sim 75\%$. Such a drop
of the indistinguishability is clear considering that the observed
field is made of the quantum superposition of two fields: one from
each 2LS. Such a superposition, together with the quantum
interference that inherently takes place in the dynamics of the
photon flux, is what makes the photons more
distinguishable. Conversely, considering the photons emitted by the
target 2LS, we find that they are more indistinguishable that those
emitted by the source 2LS: for~$\pi$-pulses, we find a $\sim 95\%$
indistinguishability. Further, although the photons from the target
2LS are always more indistinguishable than the ones from the source
2LS, the enhancement is more prominent near even-$\pi$ pulses, where
the target 2LS provides photons up to $45\%$ more indistinguishable
than the source 2LS. Therefore, quantum excitation provides an
enhancement of \emph{both} the antibunching and the
indistinguishability of the emission.

\subsection{Signatures of stimulated emission}

In the previous two subsections we have considered the correlations
between photons in different pulses. However, the question about the
correlations between photons within a single pulse remains
open. Here we provide an answer, and we will begin with the
two-particle correlation function
$G^{(2)}_c (t_1, t_2) = \mean{\ud{c}(t_2)\ud{c}(t_1) c(t_1) c(t_2)}$
with $c=\lbrace \sigma,\xi, J_\phi \rbrace$ for the source 2LS, the
target 2LS and the photon flux, respectively. Such a function
captures the correlations between photons (not necessarily
consecutive) emitted at times~$t_1$ and~$t_2$, respectively. Thus,
we compute a two-photon correlation map that, instead of their
frequencies~\cite{delvalle2012a, pietka2013, gonzalez-tudela2013,
  peiris2015, silva2016, lopezcarreno2017, lopezcarreno2018}, takes
into account their emission
times. \textbf{Figure~\ref{fig:TueNov5104935CET2024}} shows the
delayed correlations of photon pairs emitted from the source 2LS,
the target 2LS and the photon flux. Although the correlation map
from the target 2LS seems to be a copy from the correlations for the
source 2LS, the time scales in the two figures are different.  The
correlation in the target 2LS decay almost twice as slowly as their
counterpart for the source 2LS. Such an observation is also in
agreement with the fact that the emission from the target 2LS is
also more spread in time, as we showed in
Figure~\ref{fig:ThuApr10115912CEST2025} and discussed in the
previous sections. On the other hand, the correlations from the
photon flux are qualitatively different from those of the
\emph{bare} 2LSs. In fact, while the correlations
$G^{(2)}_\sigma (t,t)$ and $G^{(2)}_\xi (t,t)$ are exactly zero at
all times, because the operators $\sigma$ and~$\xi$ are nilpotent
(i.e., $\sigma^2=0$), for the photon flux we
find~$G^{(2)}_\phi(t,t) = 4 \chi_1\chi_2 \gamma_\sigma \gamma_\xi
\mean{\ud{\sigma}\sigma \ud{\xi}\xi}(t)$, which can be nonzero,
indicating the joint occupation of the two 2LS. Indeed, instead of
the fairly straight correlations lines at
either~$t_1\gamma_\sigma \approx 0$ or~$t_2\gamma_\sigma \approx 0$,
we observe the appearance of correlations along the diagonal of the
figure, that is, for~$t_1=t_2$. These correlations imply that there
are pairs of photons emitted simultaneously, which is a signature of
\emph{stimulated emission} between a single (excited) emitter and a
single photon~\cite{rephaeli2012}. It is important to keep in mind
that the diagonal line appearing on
Figure~\ref{fig:TueNov5104935CET2024}(c), which indicates emission
at zero delay, truly indicate simultaneous emission, and it is not
equivalent to the results shown in
Figure~\ref{fig:ThuApr10115912CEST2025}, where the zero delay refers
to emission within a single pulse. Furthermore, note that the
results shown in Figure~\ref{fig:TueNov5104935CET2024} correspond to
an excitation pulse of area~$\pi$, which usually is regarded as a
\emph{single photon} excitation pulse. Thus, our results also reveal
the Poissonian character of the laser pulse: sometimes, a photon
from the pulse excites the 2LS. Later, a subsequent photon from the
same pulse, which is only there because of the fluctuations of the
photon number in the coherent state of the laser, reaches the
now-excited-2LS, which then leads to the generation of two photons
in the photon flux of the system. These same-time correlations have
a dynamics similar to the one of the occupations, showing traces of
the Rabi oscillations as a function of the area of the pulse
(cf. the Supporting Information for the details). However, all of
them decay with an exponent~$\gamma_d = 2 \gamma_\sigma$, which
indicates that the lifetime of these correlations is half of the
lifetime of the spontaneous emission, i.e., $\tau_d= \tau_\sigma/2$,
in agreement with the recent experimental
observations~\cite{hansen2024}.

\begin{figure}[b]
  \begin{center}
    \includegraphics[width=\linewidth]{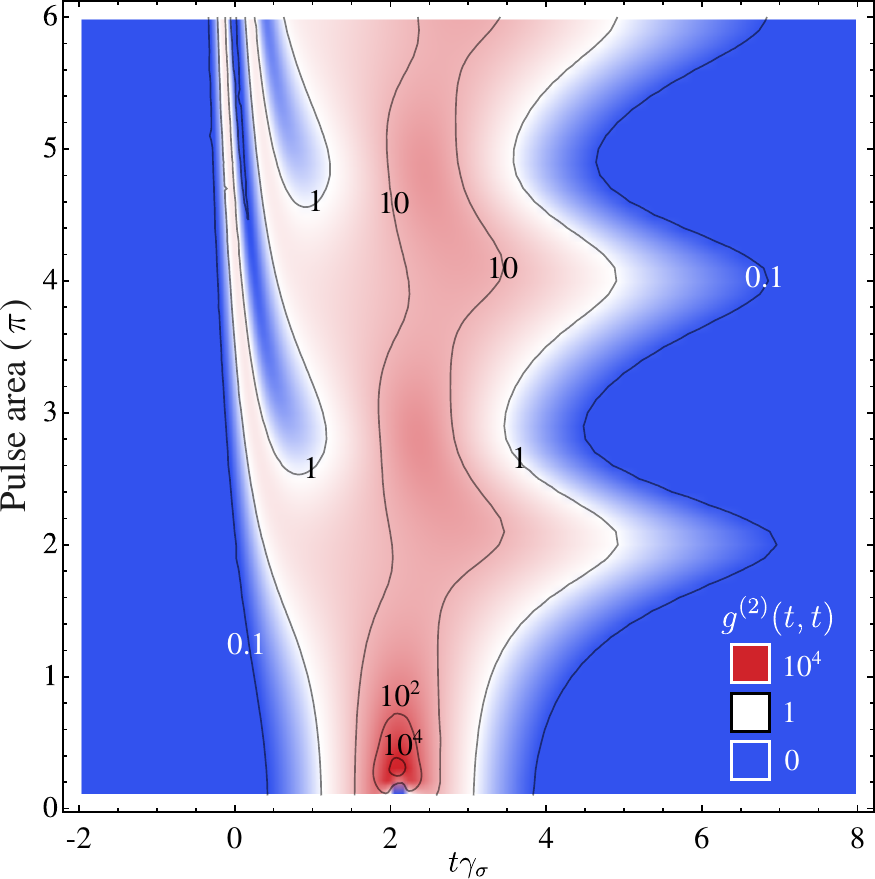}
  \end{center}
  \caption{\textbf{Normalized time-dependent correlation function of
      the photons within a single pulse, $g^{(2)}(t,t)$}. While the
    correlations of the source and target 2LS are exactly zero for
    all~$t$, the photon flux display bunching lasting for over two
    lifetimes. The figure was obtained for a pulse of
    width~$\mathcal{W}\gamma_\sigma = 1$ and letting~$\chi_2 =
    1/2$. The origin of time is set by the position of the maximum
    of the classical pulse.}
  \label{fig:TueNov5104957CET2024}
\end{figure}

Now that we have identified the quantum correlations associated to
the stimulated emission, we can quantify them through their
normalised version, shown in
\textbf{Figure~\ref{fig:TueNov5104957CET2024}}. The first notable
observation from this figure is that the process of stimulated
emission does not take place only at large pulse areas. In fact, the
photon flux always displays such a process, and it is precisely when
the classical pulses are weak that the correlations are the
strongest!  In fact, in the same way in which it was found in
Resonance Fluorescence upon frequency-filtering, the strongest
correlations appear in the regions where the emission is the
scarcest~\cite{gonzalez-tudela2013, lopezcarreno2017}. Namely, in a
process akin to distillation, the strong correlations are what is
left once the other processes are not present. The second
observation from this figure is the fact that the bunching in the
zero-delay correlations of the photon flux coincide with the
appearance of the second excitation peak in the time-dependent
occupation, which appear after the dip due to the quantum
interference of the fields, as we showed in
Figure~\ref{fig:ThuApr10115912CEST2025}. These peaks, that extends
for over two lifetimes, and for even-$\pi$ pulses are even longer,
remain bunched throughout, and therefore constitute another
mechanism to extract highly correlated photons by filtering the
emission in time.

\section{Discussion and Conclusions}
\label{sec:FriApr4114458CEST2025}

Based on recent experimental results~\cite{hansen2024}, we have
presented a thorough theoretical description of the dynamics of a
quantum system under pulsed quantum excitation.  In our description,
we have used the most fundamental, and phenomenologically rich,
quantum system available, a 2LS, both as the source and the optical
target of the quantum light. Notably, our model does not require the
addition of any dephasing mechanisms to fully reproduce the
experimental observation of our system~\cite{hansen2024}. Thus, this
indicates that for high-quality semiconductor quantum dots, the
naturally occurring dephasing processes do not spoil our
observations.  Firstly, we showed that the expected Rabi
oscillations in the total emission, predicted from the pulse area
theorem, are visible only when the length of the driving pulse is
significantly shorter than the lifetime of the 2LS. When such a
condition is not met, the visibility of the oscillations start to
decrease down to the point in which one cannot longer speak of
oscillations, as the intensity of the emission increases
monotonically. Furthermore, we found that the maxima and minima of
the Rabi oscillations are given exactly by even- and
odd-$\pi$-pulses, respectively. However, such an expectation is only
met in the limit of ultra-short pulses, and the position of the
maxima and minima drift away as the length of the pulse
increases. Thus we showed that, together with dissipation and
coupling to phonon fields, the length of the pulse is yet another
mechanism that induces a shift in the position of the maxima and
minima of the Rabi oscillations.

Continuing with the single-particle observables, we found that both
the time-integrated spectrum and the occupation the source 2LS are
inherited closely by the photon flux. On the other hand, although
the emission spectra of the target 2LS appears to be a single line,
the dynamical dressing of the source 2LS is also inherited here, but
with a much smaller intensity. However, it is noteworthy that the
linewidth of the target 2LS, driven with quantum light, is narrower
than the linewidth of the source 2LS.  In fact, a closer inspection
of the occupation of the target 2LS and the photon flux, revealed
that a considerable part of the light is emitted well beyond the
limits of one lifetime of the 2LS.

The relevance of such a delayed emission is unveiled when
considering the two-particles observables, namely the second-order
coherence function. The photon flux displays strong bunching in its
zero-delay correlation, which is an indication of stimulated
emission from the system. Such an observation is supported by two
facts: firstly they are not present in the figures of the source nor
the target 2LSs, and secondly the lifetime of these correlations is
exactly half the lifetime of the spontaneously emitted
photons. Finally, we also found that the light emitted from the
target 2LS is always more antibunched and more indistinguishable
than the emission from the source 2LS. Thus, we demonstrated that
pulsed quantum excitation can be used as a mechanism to produce
subnatural linewidth antibunched light, as shown in
Ref.~\cite{lopezcarreno2016a, lopezcarreno2018a} for continuous
coherent driving.

The approach used here, namely, the consideration of pulsed rather
than continuous excitation, makes our findings accessible to
experimental groups with state-of-the-art setups. Thus, we believe
that our results can be used as a guideline to the exploration of
quantum phenomena arising in condensed matter physics, and to work
towards the resolution of both fundamental and practical
questions. For instance, the photon flux emission is a natural
resource that can be used to exploit quantum interferences and
develop, e.g., mechanisms akin to homodyne detection, or two-photon
interference protocols, and even extract varying types of photon
correlations by selecting the frequency range of their
emission~\cite{bermudezfeijoo2025}. Alternatively, starting from the
results form the target 2LS, we can envisage a mechanism to generate
\emph{more antibunched and more indisntiguishable} photons emitted
with narrower line.  Notably, our results are also valid in regime
of operation of modern sources of single photons~\cite{somaschi2016,
  tomm2021}, namely when the length of the laser pulse is about two
orders of magnitude shorter than the lifetime of the emitter
(cf. Fig.~(S4) of the Supporting Information). In fact, such a
mechanism could be constructed in a cascaded scheme, and then one
should consider whether the antibunching, the indistinguishability
and the linenarrowing of the emission can be further improved as the
number of steps in the cascade increases. Furthermore, a natural
continuation for this work would be to include
chirping~\cite{kappe2024} into the description of the pulse, or to
consider other types of quantum light as the source of
excitation. For instance, photon streams with a varying coherence
function (which could be used as an spectroscopy tool) or other
types of quantum correlations, including entanglement.





\section*{Acknowledgements}

We thank Lena M. Hansen and Í. Arrazola for interesting discussions
and feedback on early versions of the manuscript. This research was
funded in whole by the Polish National Science Center (NCN)
“Sonatina” project CARAMEL with number 2021/40/C/ST2/00155. For the
purpose of Open Access, the authors have applied CC-BY public
copyright licence to any Author Accepted Manuscript (AAM) version
arising from this submission.





\bibliographystyle{wileyNJD-APS}
\bibliography{Sci-Camilo,Books,arXiv}

\end{document}